# Self-assembling of Ge quantum dots in an alumina matrix

M. Buljan,1 S. R. C. Pinto,2 A. G. Rolo,2 J. Martín-Sánchez,2 M. J. M. Gomes,2 J. Grenzer,3 A. Mücklich,3 S. Bernstorff,4 and V. Holý5

1 Ruđer Bošković Institute, Bijenička cesta 54, 10000 Zagreb, Croatia
2 Centre of Physics and Department of Physics, University of Minho, Campus de Gualtar, 4710-057 Braga, Portugal
3 Forschungszentrum Dresden-Rossendorf, e.V., P.O. Box 510119, 01314 Dresden, Germany
4 Sincrotrone Trieste, SS 14 km163, 5, 34012 Basovizza, Italy
5 Charles University in Prague, Ke Karlovu 5, 121 16 Prague, Czech Republic

In this work we report on a self-assembled growth of a Ge quantum dot lattice in a single 600-nm-thick Ge+Al2O3 layer during magnetron sputtering deposition of a Ge+Al2O3 mixture at an elevated substrate temperature. The self-assembly results in the formation of a well-ordered three-dimensional body-centered tetragonal quantum dot lattice within the whole deposited volume. The quantum dots formed are very small in size less than 4.0 nm, have a narrow size distribution and a large packing density. The parameters of the quantum dot lattice can be tuned by changing the deposition parameters. The self-ordering of the quantum dots is explained by diffusion-mediated nucleation and surface-morphology effects and simulated by a kinetic Monte Carlo model.

## I. INTRODUCTION

Semiconductor quantum dots (QDs) have been widely investigated in the last years because of their interesting physical properties and great potential for technological applications.1–6 Regularly ordered QDs are of a special interest, since a spatial regularity implies a narrowing of the QDs size distribution, which is very important for more pronounced quantum confinement and collective behavior effects and consequently for their potential applications.6–9 Germanium QDs embedded in amorphous wide-band-gap matrices like SiO2 or Al2O3 have numerous interesting properties such as very strong quantum confinement, electroluminescence and photoluminescence, nonlinear refraction index, possibility to retain electric charge for long time, etc.10–16 Therefore they have great possibilities for application in nanotechnology, especially for QD-based memories, sensors and solar cells. Recent investigations showed that Al2O3 matrix has many advantages compared to usually used fused silica, since alumina has a higher dielectric constant, excellent thermal and mechanical properties and it is more suitable as a building material for gates in memory devices.17 Therefore the production of regularly ordered Ge QDs in an alumina matrix is of great interest for production of new materials. While self-ordering growth of Ge QDs in silica matrix was reported recently,7,8,18,19 no similar research was carried out for alumina. Another important feature worth to note is that the self-ordering growth of QD lattice was achieved so far only by a multilayer deposition in crystalline and amorphous systems6,7 while similar observations have not been found in continuous thicker layers. Here we present a study of a self-assembled growth of Ge QDs in an alumina matrix during a continuous deposition of a Ge+Al2O3 mixture creating a single layer with approximate thickness of 600 nm. The result is the formation of a large and well-ordered three-dimensional 3D QD lattice with a body-centered tetragonal bct arrangement of QDs.
Tuning the deposition parameters it is possible to manipulate the QD sizes and the parameters of the QD lattice. The sizes of the formed QDs are homogeneous and their spatial density may be very large due their very small sizes and distances. The driving force for the observed self-ordering is explained by surface-morphology effects, i.e., by a combination of a diffusion-mediated nucleation and an enhanced probability of nucleation in surface minima. As we show later, the properties of the self-ordering of Ge QDs in alumina are different from those in silica.

The ordering in alumina is observed in a continuous layer, in contrast to silica, where the ordering can be achieved only during a multilayer growth. Another advantage of the self-assembly in alumina is that the regularity of QD positions is either not dependent on layer thickness or improves with increase of thickness, while it becomes worse during the growth of a Ge/SiO2 multilayer. In this paper we present the results of our investigation of the structure of QD lattices formed at different deposition parameters by transmission electron microscopy (TEM), atomic force microscopy (AFM), and grazing incidence small angle x-ray scattering (GISAXS). We studied also the composition and inner structure of the QDs by energy dispersive x-ray spectroscopy (EDXS), Raman scattering and x-ray diffraction (XRD) techniques. The obtained results and the type of formed QD lattices are successfully explained and simulated by a kinetic Monte Carlo model.

II. GROWTH AND ARRANGEMENT OF SELF-ASSEMBLED QUANTUM DOTS

Thin layers of a Ge+Al2O3 mixture were deposited by RF-magnetron sputtering on Si(111) substrates under elevated substrate temperatures. The sputtering target was a composite of Al2O3 (99.99%, 50 mm diameter) with pieces of polycrystalline Ge (99.99%) on top of it, covering 8±1% of the target total area. No additional oxygen flow was used.

| Sample | A | B | C |
|---|---|---|---|
| $T_d$ (°C) | 500 | 500 | 250 |
| $P_{Ar}$ (mbar) | $5 \times 10^{-3}$ | $4 \times 10^{-3}$ | $4 \times 10^{-3}$ |
| $P_{RF}$ (Watt) | 50 | 50 | 50 |
| G (nm/min) | 2.44±0.02 | 2.58±0.04 | 3.25±0.04 |
| a (nm) | 9.8±0.2 | 8.5±0.2 | 6.2±0.2 |
| c (nm) | 11.4±0.2 | 16.4±0.2 | 7.8±0.2 |
| $R_L$ (nm) | 2.2±0.2 | 1.8±0.2 | 0.7±0.1 |
| $R_V$ (nm) | 2.5±0.2 | 4.1±0.2 | 2.1±0.1 |
| $R_{cryst}$ (nm) | 1.8±0.1 | 1.1±0.1 | 0.7±0.1 |
| F (%) | 9.3±0.9 | 9.4±0.8 | 2.9±0.4 |
| $n_{QD}$ (m$^{-3}$) | $1.8 \times 10^{24}$ | $1.7 \times 10^{24}$ | $6.7 \times 10^{24}$ |

TABLE I. Deposition conditions and QD lattice parameters of the investigated samples: deposition temperature ($T_d$), Ar pressure ($p_{Ar}$), RF power ($P_{RF}$), deposition rate (G), parameters of QD lattice (a,c), QD lateral and vertical radii ($R_L$ and $R_V$), nanocrystal radius ($R_{cryst}$), filling factor F= of the quantum dots, and the QD density nQD.

Prior to sputtering, a pressure of at least $1310^{-6}$ mbar was reached inside the chamber and *in situ* argon plasma treatment of the targets and the substrates was performed in order to clean the surfaces. Different substrate temperatures ($T_d$=250 °C and 500 °C determined by the accuracy of about 10 °C and 40 °C, respectively) and different Ar pressures ($p_{Ar}$=4310$^{-3}$ mbar and 5310$^{-3}$ mbar) were used during the deposition. A larger number of samples with a broad range of deposition parameters were prepared.

Here we discuss three typical samples denoted A, B, and C; their deposition parameters are shown in Table I. The deposition rates found for different argon pressures [$p_{Ar}$(A)=5310$^{-3}$ mbar and $p_{Ar}$(B)=4310$^{-3}$ mbar] are in a good agreement with the existing literature data; the expected ratio of the growth rates $G_{A,B}$ for the Ar pressures $p_{Ar}$(A,B) is about $G_B:G_A$=1.08 (see Ref. 20) while the experimentally found value is 1.06±0.04. The filling factors F (the relative volume of the Ge nanocrystals in the layer) are found to be approximately the same for the both Ar pressures and $T_d$=500 °C (see the next sections and Table I), what indicates that the deposition rates of both Ge and Al2O3 increased in the same way with the decrease in Ar pressure.

The TEM measurements were performed using a Titan 80-300 electron microscope (FEI) equipped with a field emission gun working at 300 kV and an image corrector to minimize spherical aberrations. For AFM we used a Nanoscope III system working in tapping mode. GISAXS and XRD were measured at synchrotron Elettra (Trieste, Italy) using a photon energy of 8 keV, an image plate and a linear gas detector, respectively. In GISAXS, the two-dimensional (2D) image-plate detector was placed perpendicular to the probing sample and almost perpendicular to the incoming x-ray beam. The scattered radiation was collected for a constant incidence angle slightly above the critical angle of total reflection of the investigated films. The Raman spectra were obtained using a Jobin-Yvon T64000 with an optical microanalysis system and a charge coupled device detector, in a backscattering geometry. The Raman spectroscopy was performed at room temperature using the 514.5 nm line of an Ar laser.

The results of the structure investigations of the films are presented in Fig. 1. The formation of a regularly ordered QD system is visible in the TEM cross section of the film, [Fig. 1(a)], and it is achieved in the whole deposited volume. A very regular ordering follows also from the Fourier transformation (FT) of the TEM cross section shown in the inset of Fig. 1(a).

The enlarged images of the QD arrangements in the film cross section and on the film surface are shown in Figs. 1(b) and 1(c), respectively. From the shown cross section it follows that the QDs follow an ABAB stacking sequence while the ordering on the film surface results in a 2D square dot lattice. From this arrangement it follows that the resulting structure is a 3D bct QD lattice with the primitive vectors $a1=(a,0,0)$, $a2=(0,a,0)$ (in the plane parallel to the substrate), and $a3=(a/2,a/2,c/2)$, where $c/2=T$, $T$ is the distance between subsequent QD square layers (arrays). Thus, the formed QD lattice consists of arrays of QDs which are approximately parallel to the substrate surface and have the ABAB stacking sequence. Inside the QD arrays the QDs are arranged in 2D square lattice. The elementary unit cell with the volume $a^2c$ contains two quantum dots. From the presented TEM measurements it is obvious that the QDs are very uniform in size with the mean diameter of about 4 nm. Each QD seems to be very close or even in touch with its nearest neighbors from the arrays above and below. The regularity of ordering is very homogeneous through the whole film thickness, what is a great advantage with respect to the case of Ge/SiO2 multilayers, where only first 5–10 dot arrays in the multilayer stack are well ordered.[7]

The same ordering and size properties follow from the reciprocal-space distribution of the scattered x-ray intensity obtained by the GISAXS method, shown in Fig. 1(d). Well-pronounced Bragg spots which appear symmetrically to the $Qy=0$ plane are visible in the map, confirming the formation of a relatively well-ordered three-dimensional array of quantum dots. The positions of the Bragg peaks are the same for different azimuthal orientations of the film with respect to the probing x-ray beam, showing that the regular ordering appears in domains randomly rotated around the normal to the substrate surface.

Figures 1(e) and 1(f) present GISAXS intensity maps of films deposited under different conditions (see Table I). All the maps demonstrate the same type of the QD arrangement following from the same type of the arrangement of Bragg peaks in the maps. However, the positions of the peaks in the $QyQz$ space are different, thus, the QDs are ordered in the same lattice type for all films, but the parameters of the formed QD lattices depend on the deposition conditions. We have analyzed numerically the measured GISAXS maps assuming arrangement of QDs in a bct paracrystal lattice and an ellipsoidal shape of QDs s$RL$ and $RV$ are the dot radii parallel and perpendicular to the sample surface, respectivelyd; the model used for the fitting is presented in Ref. 7. From the fits we have determined the QD lattice parameters $a$ and $c$ and QD size parameters $RL$ and $RV$. The results are summarized in Table I.

From the fitting results it follows that the parameters of the QD lattice and the QD sizes are very sensitive to the deposition conditions; the in-plane lattice parameter $a$ can be varied in the range from 6 to 10 nm while $c$ ranges from 8 to 16 nm. The in-plane lattice parameters of films A and B (both being deposited at 500 °C) have values that are considerably larger than the same parameter of film C (deposited at 250 °C). This finding demonstrates that the lateral dot distance is mainly governed by the deposition temperature, which is consistent with the temperature dependence of the diffusion length (via temperature dependence of diffusion constant).[21] A smaller diffusion length results in a smaller lateral dimension of the dots and in the smaller lateral dot distance. On the other hand, the diffusion length depends also on the diffusion time, which is determined by the growth rate and by the total duration of the growth. The growth rate is smaller for film A than for film B so that the diffusion time is longer for film A and consequently larger lateral values of QD sizes and QD distances can be found in this sample. The distance $T$ between the QD arrays is found increasing with the decrease in the Ar pressure for the films deposited at the same temperature (500 °C). This observation is also in a good agreement with the expectation of the formation of smaller dots at higher Ar pressure.[21]

The shape of the QDs is found to be elongated in the vertical direction (or more precisely in direction of the primitive vector $a_3$). The QD lateral and vertical radii $R_L, R_V$ can be tuned in the ranges of 0.7–2.2 nm and 2.1–4.1 nm, respectively. From the comparison of the distance $T$ of the QD 2D arrays with the QD radius $R_V$ it follows that $T \sim 2R_V$, i.e., the distance between QD arrays is approximately equal to the vertical size of the QDs. The same arrangement is well visible in TEM cross sections of the film [Figs. 1(a) and 1(b)].
The sample C is of a particular interest, since it contains very small and densely packed QDs. This sample was deposited under low temperature and Ar pressure of $4 \times 10^{-3}$ mbar, the lateral dot diameter is 1.4 nm and the dot density calculated from the lattice parameters $a, c$ is $n_{QD} = 6.7 \times 10^{24}$ QDs/m$^3$.

III. COMPOSITION AND CRYSTALLINE STRUCTURE OF THE FILMS

For the investigation of the composition and the inner structure of the formed QDs and the surrounding matrix, EDXS, Raman and XRD measurements were used. Results of the EDXS measurements performed on sample A are shown in Fig. 2. The measurements were carried out in two areas of the cross section of the layer indicated in the inset of Fig. 2. The elemental compositions of both areas are very similar, showing the presence of significant amounts of Al, Ge, and O. The atomic percentages of these elements are found to be $c_{Ge} = (17 \pm 1)\%$, $c_{Al} = (29 \pm 1)\%$, and $c_O = (47 \pm 1)\%$ for the area 1 and $c_{Ge} = (15 \pm 1)\%$, $c_{Al} = (32 \pm 1)\%$, and $c_O = (53 \pm 1)\%$ for area 2. Very small amounts (less than 1%) of Si and Ar were found as well. The positions of maximal concentration of oxygen atoms coincide with the positions of Al atoms sfor more details see Ref. 22d, no evidence of the presence of GeO2 was found in the infrared or Raman spectra of the films. As well, the ratio of the percentages of Al and O is found to be $c_O : c_{Al} = 1.6 \pm 0.1$, i.e., very close to the value 1.5 of the ideal Al2O3 matrix. The amount of nonstoichiometric additional oxygen atoms sdenoted by the hatched areas in the bar plot in Fig. 2d is comparable to the accuracy of the EDXS methods. Therefore practically all O atoms are bonded to Al and no considerable amount of GeO2 is present in the films.

The results of Raman and XRD measurements presented in Fig. 3 confirm the formation of Ge quantum dots during the deposition in all three films. In the x-ray diffraction graphs shown in Fig. 3(a) crystalline Ge peaks are well visible in sample A, which is deposited under a high pressure and at a temperature of 500 °C. Much wider diffraction peaks are present for the other two samples, indicating the presence of smaller Ge nanocrystals in them. From the width of the diffraction peaks we have calculated the sizes of Ge coherent nanocrystals using the standard Williamson-Hall procedure assuming nearly spherical nanocrystals within the QDs; the results are presented in Table I. From the comparison of the sizes of the quantum dots obtained from GISAXS and the sizes of the nanocrystals from XRD it follows that the radius $R$ cryst of the crystalline part of the QD is approximately equal to the lateral radius of the entire QD ($R_{cryst} \sim R_L$). To clarify these results we want to point out that the GISAXS technique is sensitive only to the electrondensity contrast between the QD and the surrounding matrix, not to the crystal structure of the dots. Thus, the total size of the QDs can be estimated by GISAXS while the size of the crystalline part(s) of the dots follows from XRD. Since the vertical size $2R_V$ of the QDs is larger than the lateral one, the vertically elongated QDs could consist of several randomly rotated coherent nanocrystals with the diameter of $2R_{cryst}$.
The Raman spectra of the samples shown in Fig. 3(b) give similar results: the crystalline Ge TO peak is clearly visible in sample A while amorphouslike Ge peaks appear for the other two samples where nanocrystals are smaller than 2.5 nm. No bands close to 400 cm$^{-1}$ due to Ge-O vibrations are visible in the shown Raman spectra. Alumina matrix is found to be amorphous by both XRD and Raman techniques; the same is following from the high-resolution TEM images.

IV. MONTE CARLO SIMULATION OF THE SELFASSEMBLY PROCESS

Here we explain the observed arrangement properties of the QDs. First we want to emphasize that if we deposit only the alumina matrix, no GISAXS features are visible22 showing no local inhomogeneities in alumina electron density are present. As well, no XRD maxima are measured, i.e., the alumina is fully amorphous. Therefore the matrix itself does not drive the self-ordering of QDs. The films were deposited at elevated substrate temperatures, which induced diffusionmediated nucleation and the formation of QDs during the growth of the films. We want to point out that the alumina matrix is fully amorphous, so the strain effects like the ones presented in Ref. 6 are not efficient here for the selfassembly process. As well, the arrangement of the formed QDs (i.e., the QD lattice parameters) and the QD sizes are found very dependent on the deposition conditions. Therefore, it is very likely that the ordering is driven by the surface-area minimization during the film growth, i.e., by the morphology of the growing surface. This assumption is supported also by the fact that the nucleation of Ge QDs is enhanced within the surface minima23 and by the Monte Carlo simulations that are presented in the following.

Let us discuss now the basic principles of the mentioned self-assembling growth type. In the beginning of the film deposition, the QDs form on the substrate surface due to a standard diffusion-driven homogeneous nucleation, making a 2D QD array. The growing QDs are higher than the depth of the surrounding matrix [see AFM results in Fig. 1(c)], so the surface exhibits hills above the buried QDs and troughs between them. Thus, the shape of the actually growing surface is correlated with the positions of the QDs underneath. After reaching some critical thickness, further growth of this QD array is not energetically favorable any more (the surface area is too large), so a new Ge QD array starts to grow in the troughs (minima) of the existing surface. In this way the arrangement of the QDs in the first array influences strongly the dots arrangement in the next array. The same mechanism is valid for all subsequent arrays inducing a self-assembling. The vertical distance between the centers of two QD arrays $T$ is actually equal to the vertical size of the QDs, what was experimentally found by TEM and GISAXS analyses (Table I).

A similar ordering mechanism was used for the explanation of the self-assembly in Ge/SiO2 multilayers presented in Ref. 7. In these samples Ge+SiO2 mixture layers were approximately 10 nm thick and they were separated by pure SiO2 layers. Thus, the separation of the 2D QD arrays in the Ge+SiO2 mixture layers was determined by the multilayer periodicity. The QDs in silica were arranged in a disordered rhombohedral (fcc-like) lattice, which appeared in small domains randomly rotated around the substrate normal. Similarly to the samples with the alumina matrix, an elevated substrate temperature gave rise to a QD formation during the deposition. Hills were formed as well above each buried QD lying in the previous Ge+SiO2 layer in the multilayer stack and the Ge QDs in the growing layer nucleated in the surface troughs between the QDs in the previous layer. We have shown that the two topmost dot layers influence the growing surface in the samples with silica matrix, and this fact was crucial for the achievement of a ABCABC layer stacking and fcc-like growth of the QD lattice.7

For the numerical simulation of the self-ordering of Ge quantum dots in a single Al2O3+Ge layer we use the same concept as in Ref. 7 and we investigate the time evolution of the arrangement of the formed QDs and of the morphology of the growing surface. The surface used for the deposition was Si(111), so we can assume that the first deposited layer has a distorted 2D hexagonal lattice because of the diffusion kinetics (Voronoi cells) on the isotropic surfaces.7,24 The same substrate was used for the Ge+SiO2 systems so that the ordering of the Ge QDs close to the substrate is similar, however, the stacking sequence in the case of alumina matrix (ABAB or bcc-like) is different than for the silica matrix (ABCABC or fcc-like). We have found experimentally that the hills above the buried array of QDs are smaller in alumina than in the silica. That is visible in the TEM cross sections of the films: surface is smoother, i.e., the hills above the QDs are smaller in the alumina matrix. Therefore we can assume that the influence of the dot arrays buried under the growing array is weaker in alumina than in the silica matrix, and only the topmost dot array can influence the surface shape in the latter case. Using the Monte Carlo simulation algorithm published in Ref. 7, we found the simulation parameters to obtain the agreement with the experimentally found QD arrangement. As it is described in the mentioned reference, we assume that a Gaussian-type hill forms above each QD. The shape of the surface at the interface $j_3$ depends on the arrangement of the QDs just below the current surface (formed at interface $j_3$-1) and in one array deeper (interface $j_3$-2). We describe the morphology of the $j_3$th interface by the following formula:

$$h_{j_3}(x) = \sum_{j_1, j_2} [f(x - X_{j_1, j_2, j_3-1}) + Cf(x - X_{j_1, j_2, j_3-2})],$$

$$f(x) = \exp(-|x|^2/b^2), \quad x = (x, y),$$

where $f(x)$ is the function determining the shape of the hill above a buried dot (assumed Gaussian for simplicity), $b$ is a parameter defining the width of the formed hills, and $C$ expresses the influence of the array $j_3$-2 on the morphology of the surface of array $j_3$. A good agreement with the experimentally observed structure was obtained for a much smaller value of the parameter $C$ than for the SiO2 matrix so that the array $j_3$-2 does not influence significantly the morphology of the top array. This result is indeed consistent with experimental finding that the alumina matrix grows smoother than silica for the deposition conditions given here (alumina) and in Ref. 7 (silica).

The results of the simulations are shown in Fig. 4. In the simulation we have assumed that the QDs grow in 2D QD arrays parallel to the substrate surface what is consistent with TEM and AFM experimental findings. The vertical distance of the QD arrays corresponds to the vertical size of the QDs (this is a reasonable assumption if the formed QDs have similar sizes). When one array is completed the next array starts to grow. The simulated arrangements of the QDs in the planes parallel to the surface are shown in Fig. 4(a). From these simulations it follows that the in-plane ordering of QDs changes from a distorted 2D hexagonal lattice, which is assumed in the first array, to distorted squared lattices in the following arrays. The cross-sectional view of the simulated QD arrangement is shown in Fig. 4(b). The ordering depicted here has the same parameters as that found in the TEM cross sections and AFM images of the film surface.

The explanation of the transformation of the in-plane QD arrangement from a hexagonal to a square 2D lattice is schematically shown in Fig. 4(c). Let us follow the positions of the surface minima when one array is already grown. We assume that Gaussian-type hills form above each QD formed. If the disorder in the starting array is small, six well-separated minima form above each QD (left panel) while for the case of a higher disorder some of the minima will be very close or overlap (right panel). In the latter case the coordination number of the dots is reduced since only one QD can grow in these overlapping (or very close) minima. If the influence of all layers (more than one) below the growing array is small, then the in-plane coordination reduces from six to four very rapidly and the resulting in-plane structure is a 2D square lattice. When the square lattice is formed, a new QD layer is placed in the minima of it, so it has the same square arrangement of QDs. The only possible QD layer stacking is ABAB if the most significant contribution to the surface morphology comes from the last layer. The resulting structure for this growth mechanism is the 3D bct lattice, what is observed experimentally indeed. Such a distinct squarelike ordering would not be possible for a larger value of the parameter C. For a larger C, i.e., if the array $j_3$-2 more affects the shape of the growing surface, the fcc-like stacking with a hexagonal base will form.

The formed bct QD lattice is more stable than the one with fcc-like ordering observed in silica because the number of minima at the surface for 2D square lattice equals the number of QDs buried in the layer below (for hexagonal lattice there are two times more minima), and only one layer actually influences the QD ordering. Therefore, the regularity in QD ordering is easily achieved and improving with the number of deposited layers. In a fcc-like ordering, the stacking of subsequent dot arrays can easily change from ABCABC to ABAB and vice versa, which deteriorates the structural quality of the resulting 3D dot crystal. This observation is also well supported experimentally because the regularity of ordering in alumina is the same or even improving with the number of the deposited layers while for the silica case the regularity is decreasing. Summarizing, the Monte Carlo simulations agree very well with the all main experimentally observed properties of QD lattices in alumina matrix.

V. SUMMARY

In summary, we have demonstrated formation of large and well-ordered 3D QD lattices of Ge QDs in amorphous alumina matrix by a self-assembly process. The observed QD self-assembling is explained by nucleation during the deposition combined with surface-morphology effects and simulated by a Monte Carlo model. In contrast to more common quantum dots in silica matrix, the formed QD lattices have bct QD packing and their lattice parameters depend on the deposition conditions. The formed QDs have sizes of about 4 nm, a narrow size distribution, and a very dense QD packing ($6.7 \times 10^{24}$ QDs/m$^3$). The observed self-assembly mechanism represents a distinct route for the creation of ordered QD arrays in alumina matrices.

ACKNOWLEDGMENTS

This work has been partially funded through the Project PTDC/FIS/70194/2006 financed by the Portuguese Foundation for Science and Technology (FCT). The authors thank ELETTRA and the European Community for financial support under EU Contract No. RII3-CT-2004-506008 for the measurements at SAXS beamline (Proposal 2008345). S.R.C.P. and J.M.S. are grateful for the FCT under Grants No. SFRH/BD/29657/2006 and SFRH/BPD/64850/2009, respectively. M.B. acknowledges support from the Ministry of Science Education and Sports, Republic Croatia (Project No. 098-0982886-2866). M.B. and V.H. acknowledge the NAMASTE EU project (Contract No. 214499). We would like to thank S. Levichev for his help with the measurements at ELETTRA and P. Caldelas for the sample preparation.

References:

1 R. Hanson, L. P. Kouwenhoven, J. R. Petta, S. Tarucha, and L. M. K. Vandersypen, Rev. Mod. Phys. 79, 1217 (2007).
2 F. H. Julien and A. Alexandrou, Science 282, 1429 (1998).
3 T. C. Chang, S. T. Yan, P. T. Liu, C. W. Chen, S. H. Lin, and S. M. Sze, Electrochem. Solid-State Lett. 7, G17 (2004).
4 J. Konle, H. Presting, H. Kibbel, K. Thinke, and R. Sauer, SolidState Electron. 45, 1921 s2001d.
5 J. J. Lee, X. G. Wang, W. P. Bai, N. Lu, and D.-L. Kwong, IEEE Trans. Electron Devices 50, 2067 (2003).
6 J. Stangl, V. Holý, and G. Bauer, Rev. Mod. Phys. 76, 725 (2004).
7 M. Buljan, U. V. Desnica, M. Ivanda, N. Radić, P. Dubček, G. Dražić, K. Salamon, S. Bernstorff, and V. Holý, Phys. Rev. B 79, 035310 (2009).
8 M. Buljan, U. V. Desnica, M. Ivanda, N. Radić, P. Dubček, G. Dražić, K. Salamon, S. Bernstorff, and V. Holý, Nanotechnology 20, 085612 (2009).
9 D. Grützmacher, T. Fromherz, C. Dais, J. Stangl, E. Müller, Y. Ekinci, H. H. Solak, H. Sigg, R. T. Lechner, E. Wintersberger, S. Birner, V. Holý, and G. Bauer, Nano Lett. 7, 3150 (2007).
10 S. K. Ray and K. Das, Opt. Mater. sAmsterdam, Neth.d 27, 948 (2005).
11 A. Dowd, R. G. Elliman, M. Samoc, and B. Luther-Davies, Appl. Phys. Lett. **74**, 239 (1999).

12 M. Kanoun, C. Busseret, A. Poncet, A. Souifi, T. Baron, and E. Gautier, Solid-State Electron. **50**, 1310 (2006).
13 S. H. Hong, M. C. Kim, P. S. Jeong, S. H. Choi, and K. J. Kim, Nanotechnology **19**, 305203 (2008).
14 C. W. Teng, J. F. Muth, R. M. Kolbas, K. M. Hassan, A. K. Sharma, A. Kvit, and J. Narayan, Appl. Phys. Lett. **76**, 43 (2000).
15 P. Tognini, L. C. Andreani, M. Geddo, A. Stella, P. Cheyssac, R. Kofman, and A. Migliori, Phys. Rev. B **53**, 6992 (1996).
16 J. Narayan, J. Nanopart. Res. **2**, 91 (2000).
17 E. P. Gusev, E. Cartier, D. A. Buchanan, M. Gribelyuk, M. Copel, H. Okorn-Schmidt, and C. D'Emic, Microelectron. Eng. **59**, 341 (2001).
18 M. Buljan, I. Bogdanović-Radović, M. Karlušić, U. V. Desnica, G. Dražić, N. Radić, P. Dubček, K. Salamon, S. Bernstorff, and V. Holý, Appl. Phys. Lett. **95**, 063104 (2009).
19 M. Buljan, I. Bogdanović-Radović, M. Karlušić, U. V. Desnica, N. Radić, N. Skukan, G. Dražić, M. Ivanda, O. Gamulin, Z. Matej, V. Valeš, J. Grenzer, T. W. Cornelius, H. T. Metzger, and V. Holý, Phys. Rev. B **81**, 085321 (2010).
20 T. P. Drüsedau, M. Löhmann, and B. Garke, J. Vac. Sci. Technol. A **16**, 2728 (1998).
21 I. Petrov, P. B. Barna, L. Hultman, and J. E. Greene, J. Vac. Sci. Technol. A **21**, S117 (2003).
22 S. R. C. Pinto, A. G. Rolo, M. J. M. Gomes, M. Ivanda, I. Bogdanović-Radović, J. Grenzer, A. Mücklich, D. Barber, S. Bernstorff, and M. Buljan, Appl. Phys. Lett. **97**, 173113 (2010)
23 A. Karmous, I. Berbezier, and A. Ronda, Phys. Rev. B **73**, 075323 (2006).
24 F. Ratto, A. Locatelli, S. Fontana, S. Kharrazi, S. Ashtaputre, S. K. Kulkarni, S. Heun, and F. Rosei, Phys. Rev. Lett. **96**, 096103 (2006).

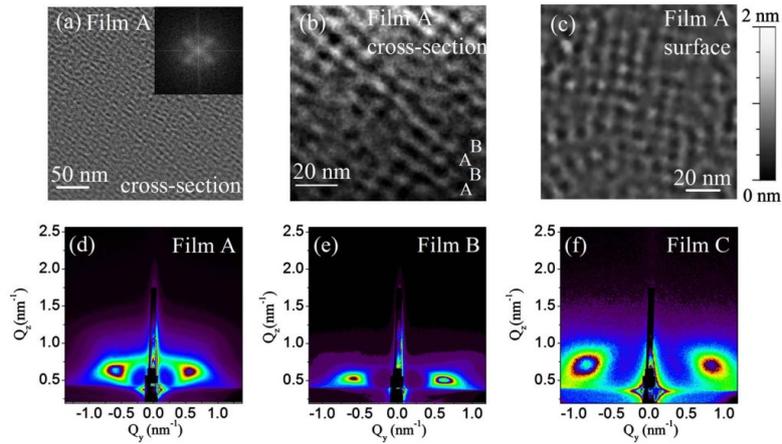

FIG. 1. Structure of the grown films. (a) TEM cross section of film A. The substrate surface is parallel to the bottom edge of the shown cross section. The inset shows the corresponding FT, (b) TEM cross section of film A with a better resolution, showing an ABAB type of QD ordering. (c) AFM image of a film A surface showing squarelike in-plane ordering of QDs. [(d)–(f)] GISAXS maps of films A–C grown under different conditions (Table I).

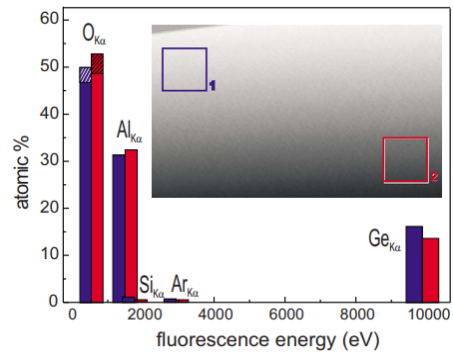

FIG. 2. The EDXS spectra of sample A. The measurements were performed in area 1 closer to the layer surface and in area 2 closer to the layer/substrate interface. The areas are indicated in the TEM cross-section image of the film shown in the inset. The hatched parts of the bars denote the amount of surplus (nonstoichiometric) oxygen.

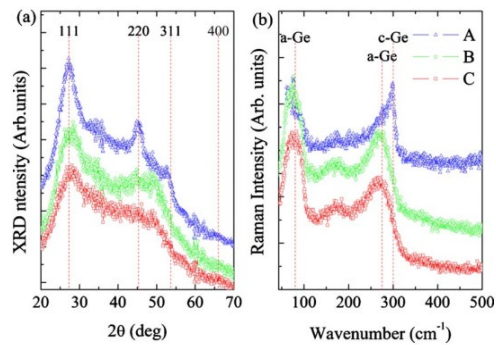

FIG. 3. (a) X-ray diffraction curves: the vertical dashed lines show the positions of the diffraction peaks for crystalline Ge. (b) Raman spectra: amorphous (a-Ge) and crystalline (c-Ge) frequency modes are indicated by vertical dashed lines.

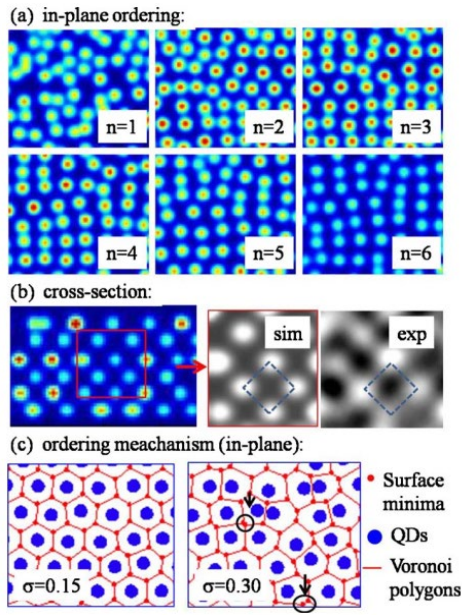

FIG. 4. Explanation and Monte Carlo simulations of the growth of a bcc-like QD lattice. (a) In-plane arrangement of QDs in various 2D dot arrays, *n* is the array index starting from the bottom of the layer. (b) Cross-sectional view of the positions of QDs. The insets show simulated (sim) and experimentally measured by TEM (exp) QD arrangements. (c) If the initial disorder in a 2D hexagonal lattice is low (left panel, σ=0.15) six well-separated minima exist at the surface around each buried dot. For the higher disorder degree (right panel, σ=0.30), some of the minima are very close or overlap (indicated by arrows).